\documentclass[traditabstract]{aa}
\usepackage{graphicx}
\usepackage{natbib}
\usepackage{amssymb}
\usepackage{amsfonts}
\usepackage{amsbsy}
\usepackage{amsmath}

\bibpunct{(}{)}{;}{a}{}{,}

\newcommand{\Scmbi}{\mathfrak{S}^{(i)}_{\rm c}(p)}
\newcommand{\Scmb}{\mathfrak{S}_{\rm c}(p)}

\newcommand{\Scbj}{\boldsymbol{\mathfrak{S}}_j}

\newcommand{\Scba}{\boldsymbol{\mathfrak{S}}_1}
\newcommand{\Scbb}{\boldsymbol{\mathfrak{S}}_2}
\newcommand{\ScbNc}{\boldsymbol{\mathfrak{S}}_{Nc}}
\newcommand{\Scmbh}{\widehat{\mathfrak{S}}_{\rm c}(p)}
\newcommand{\Scmbhb}{\widehat{\boldsymbol{\mathfrak{S}}}_{\rm c}}
\newcommand{\Scmbb}{\boldsymbol{\mathfrak{S}}_{\rm c}}

\newcommand{\Smth}{\mathfrak S}
\newcommand{\Smthb}{\boldsymbol{\mathfrak S}}

\newcommand{\Sgalb}{\boldsymbol{S}_{\rm f}}
\newcommand{\Sgalib}{\boldsymbol{S}^{(i)}_{\rm f}}
\newcommand{\Sib}{\boldsymbol{S}^{(i)}}
\newcommand{\Sgal}{S^{(i)}_{\rm f}(p)}
\newcommand{\Si}{S^{(i)}(p)}
\newcommand{\Shi}{\widehat{S}^{(i)}(p)}

\newcommand{\Nmthi}{{\mathcal N}^{(i)}}
\newcommand{\Nmthib}{\boldsymbol{{\mathcal N}}^{(i)}}
\newcommand{\Nmthb}{\boldsymbol{\mathcal N}}
\newcommand{\Fmatb}{\boldsymbol{{\mathcal F}}}

\newcommand{\oneb}{\boldsymbol{1}}
\newcommand{\Ab}{\boldsymbol{A}}
\newcommand{\Fb}{\boldsymbol{F}}

\newcommand{\Cb}{\boldsymbol{C}}

\newcommand{\Hb}{\boldsymbol{H}}
\newcommand{\Ib}{\boldsymbol{I}}

\newcommand{\Omegab}{\boldsymbol{\Omega}}

\newcommand{\Shb}{\boldsymbol{\widetilde S}}
\newcommand{\Sb}{\boldsymbol{S}}

\newcommand{\wb}{\boldsymbol{w}}
\newcommand{\zerob}{\boldsymbol{0}}

\begin{document}

   \title{``Internal Linear Combination'' method for the separation of CMB from Galactic foregrounds
    in the harmonic domain}
   \author{R. Vio\inst{1}
    \and
           P. Andreani\inst{2}
          }
   \institute{Chip Computers Consulting s.r.l., Viale Don L.~Sturzo 82,
              S.Liberale di Marcon, 30020 Venice, Italy\\
              \email{robertovio@tin.it},
         \and
		  ESO, Karl Schwarzschild strasse 2, 85748 Garching, Germany\\
                  INAF-Osservatorio Astronomico di Trieste, via Tiepolo 11, 34143 Trieste, Italy\\              
		  \email{pandrean@eso.org}
             }

\date{Received .............; accepted ................}

\abstract
{Foreground contamination is the fundamental hindrance to the cosmic microwave background (CMB) signals and its separation from it represents a fundamental
question  in Cosmology. One of the most popular algorithm used to disentangle foregrounds from the CMB signals
is the ``{{\it internal linear combination}}'' method (ILC). In its original
version, this technique is applied directly to the observed maps. In recent literature, however, it is suggested that in the harmonic (Fourier) domain 
it is possible to obtain better results since a separation can be attempted where the various Fourier frequencies are given different weights.
This is seen as a useful characteristic in the case of noisy data. Here, we argue that the benefits of using such
an approach are overestimated. Better results can be obtained if a classic procedure is adopted where data are filtered before the separation
is carried out.}
\titlerunning{ILC method}
\authorrunning{R. Vio, \& P. Andreani}
\maketitle

\section{Introduction}

The experimental progresses in the detection of cosmological emissions require a 
parallel development of data analysis techniques in order to extract the maximum physical information 
from data. In particular, different emission mechanisms are characterized by markedly distinct underlying 
physical processes. Data analysis often requires the component separation in order to study the individual characteristics.
To achieve such a goal, a link between the branch of signal processing science 
which characterizes and separates different signals and astrophysics is yet well established, and in many cases, modern signal 
processing techniques have been imported and applied in an astrophysical context. 

In this work, we consider one of the most widely used approaches for the separation of different emissions, say
the {\it internal linear combination} method (ILC), 
which was adopted, for instance, in the reduction of the data from the Wilkinson Microwave Anisotropy Probe (WMAP) satellite for CMB observations 
\citep{ben03}. Among the separation techniques, ILC requires the smallest number of {\it a priori} assumptions.
Here, the available data are assumed in the form of $N_o$ maps, taken at different frequencies, containing $N_p$ pixels each.  
More precisely, if $\Si$ provides the value of the $p$th pixel for a map obtained at channel ``$~i~$'' \footnote{In the present work, $p$ 
indexes pixels in the classic spatial domain. However, the same formalism applies if other domains are considered, for example, the Fourier one.}, 
our starting model is:
\begin{equation} \label{eq:image}
\Si = \Scmbi + \Sgal + \Nmthi(p)
\end{equation}
where $\Scmbi$, $\Sgal$ and $\Nmthi(p)$ are the contributions due to the CMB, the diffuse Galactic foreground and the experimental noise, 
respectively.  Although not necessary for later arguments, it is assumed that all of these contributions are representable by means of stationary 
random fields. Moreover, without loss of generality, for ease of notation the random fields are supposed as the
realization of zero-mean spatial processes. In the present work the contribution of non-diffuse components 
(e.g., due to SZ cluster, point-sources, \ldots) are not considered and they are assumed to have been removed through other methodologies. 

The idea behind ILC is simple. The main assumption is that model~(\ref{eq:image}) can be written as
\begin{equation} \label{eq:basicm}
\Si = \Scmb + \Sgal + \Nmthi(p),
\end{equation}
i.e. the template of the CMB component is independent of the observing channel. A  way to exploit this fact is to 
average $N_o$ images $\{ \Si \}_{i=1}^{N_o}$ giving a specific weight $w_i$ to each of them so as to minimize the 
impact of the foreground and noise
\citep{ben03}. This means to look for a solution of type
\begin{equation} \label{eq:sum}
\Scmbh = \sum_{i=1}^{N_o} w_i \Si.
\end{equation}
If the constraint $\sum_{i=1}^{N_o} w_i = 1$ is imposed, Eq.~(\ref{eq:sum}) becomes
\begin{equation} \label{eq:wls}
\Scmbh = \Scmb + \sum_{i=1}^{N_o}  w_i [ \Sgal + \Nmthi(p) ]. 
\end{equation}
Now, from this equation it is clear that, for a given pixel ``$p$'',  the only variable terms are in the sum. Hence,  
under the assumption of independence of $\Scmb$ from $\Sgal$ and $\Nmthi(p)$,
the weights $\{ w_i \}$ have to minimize the variance of $\Scmbh$, i.e.
\begin{align}
& \{ w_i \} = \underset{\{ w_i \} }{\arg\min} \nonumber \\
& {\rm VAR} \left[\Scmb \right] + {\rm VAR}\left[\sum_{i=1}^{N_o}  w_i (\Sgal + \Nmthi(p)) \right],
\end{align}
where ${\rm VAR}[s(p)]$ is the {\it expected variance} of $s(p)$.
If $\Sib$ denotes a {\bf row vector} such as $\Sib = [S^{(i)}(1), S^{(i)}(2), \ldots, S^{(i)}(N_p)]$ and the $N_o \times N_p$ matrix $\Sb$ is defined as  
\begin{equation}
\Sb = 
\left( \begin{array}{c}
\Sb^{(1)} \\
\Sb^{(2)} \\
\vdots \\
\Sb^{(N_o)}
\end{array} \right),
\end{equation}
then Eq.~(\ref{eq:basicm}) becomes
\begin{equation} \label{eq:basicmm}
\Sb = \Scmbb + \Sgalb + \Nmthb.
\end{equation}
In this case, the weights are given by \citep{eri04}
\begin{equation} \label{eq:wr}
\wb = \frac{\Cb_{\Sb}^{-1} \oneb}{\oneb^T \Cb_{\Sb}^{-1} \oneb},
\end{equation}
where 
$\Cb_{\Sb}$ is the $N_o \times N_o$ cross-covariance matrix of the random processes that generate $\Sb$, i.e.
\begin{equation}
\Cb_{\Sb} = {\rm E}[\Sb \Sb^T], 
\end{equation}
and $\oneb = (1, 1, \ldots, 1)^T$ is a column vector of all ones. Here, ${\rm E}[.]$ denotes the {\it expectation operator}.  
Hence, the ILC estimator takes the form
\begin{align} 
\Scmbhb & = \wb^T \Sb, \label{eq:wlss} \\
        & = \alpha \oneb^T \Cb_{\Sb}^{-1} \Sb, \label{eq:basic}
\end{align}
with 
$\oneb^T \wb = 1$ and the scalar quantity $\alpha$ given by
\begin{equation} \label{eq:alpha}
\alpha = [ \oneb^T \Cb_{\Sb}^{-1} \oneb]^{-1}.
\end{equation}

In practical applications, matrix $\Cb_{\Sb}$ is unknown and has to be estimated from the data. Typically, this is done by means of the
estimator
\begin{equation} \label{eq:C}
\widehat \Cb_{\Sb} = \frac{1}{N_p} \Sb \Sb^T.
\end{equation}

A common assumption in CMB observations is that $\Sgalib$ is given by the linear mixture of the contribution of 
$N_c$ physical processes $\{ \Scbj \}_{j=1}^{N_c}$ (e.g., free-free, dust re-radiation, 
\ldots)
\begin{equation} \label{eq:gal}
\Sgalib = \sum_{j=1}^{N_c} a_{ij} \Scbj,
\end{equation}
with $a_{ij}$ constant coefficients. In practice, it is assumed that for the $j$th physical process a template $\Scbj$ exists
independent of the specific channel ``$~i~$''. Although rather strong, it is not unrealistic to assume that this condition is satisfied 
when small enough patches of the sky are considered. Inserting Eq.~(\ref{eq:gal}) into Eq.~(\ref{eq:basicmm}) one obtains
\begin{equation} \label{eq:model}
\Sb = \Ab \Smthb + \Nmthb
\end{equation}
with
\begin{equation} \label{eq:modelS}
\Smthb = \left( \begin{array}{l}
\Scmbb \\
\Scba \\
\Scbb \\
\vdots \\
\ScbNc \\
\end{array} \right),
\end{equation}
and
\begin{equation}
\Ab = 
\left( \begin{array}{cccccc} \label{eq:Amatrix}
1 & \vline & a_{11} & a_{12} & \ldots & a_{1 N_c} \\
\hline
1 & \vline & a_{21} & a_{22} & \ldots & a_{2 N_c} \\
\vdots & \vline & \vdots & \vdots & \ddots & \vdots \\
1 & \vline & a_{N_o 1} & a_{N_o 2} & \ldots & a_{N_o N_c}
\end{array} \right).
\end{equation}
Here, matrix $\Ab$ is assumed to be of full rank.
 
\section{ILC in the spatial domain}
\label{sec:ILC}

In a recent paper, \citet{vio08} have stressed various problems concerning ILC that in literature have been underevaluated if not undetected.
For example, in Eq.~(\ref{eq:basicmm}) 
the term $\Sgalb + \Nmthb $ is often considered as a single noise component 
\citep[e.g., see ][]{eri04,hin07, kim07, kim08}. 
In this way the problem is apparently simplified since it is reduced to the separation of two components only. No {\it a priori} information
on this ``global'' noise is required. However, this approach can lead to wrong 
conclusions. For example, since all the components in the mixtures $\Sb$ are assumed 
to be zero-mean, from Eq.~(\ref{eq:wls}) one could conclude that
\begin{equation}
{\rm E}[\Scmbhb | \Scmbb] = \Scmbb + \wb^T {\rm E}[\Sgalb + \Nmthb] = \Scmbb,
\end{equation}
i.e. the ILC estimator is unbiased \footnote{The expression ${\rm E}[a|b]$ indicates {\it conditional expectation}
of $a$ given $b$. }. This is not correct: the claim that $\Scmbhb$ is unbiased requires one to prove that
\begin{equation}
{\rm E}[\Scmbhb | \Scmbb, \Sgalb] = \Scmbb + \wb^T \Sgalb + \wb^T {\rm E}[\Nmthb] = \Scmbb.
\end{equation}
The reason is that $\Sgalb$ is a fixed realization of a random process. There is no way to obtain another one. 
Even if observed many times (under the same experimental conditions) the foreground components (for instance the Galaxy)
will always appear the same.
Only the noise component $\Nmthb$ will change. This has important consequences. In fact,
in \citet{vio08} it is proved that, even under model~(\ref{eq:model}), ILC can provide satisfactory results only under 
rather restrictive conditions:
\begin{enumerate}
\item The number of observing frequencies $N_o$ has to be larger than the number of components $N_c$; \\
\item $\Scmbb$ has to be uncorrelated with $\Sgalb$; \\
\item Data have to be noise-free, i.e. $\Nmthb = \zerob$.
\end{enumerate}
The violation of any of these points has as consequence the introduction of a bias in the ILC solution that can be severe.
In literature it appears that only the second point has been well realized. \citep[e.g. see][]{del07}.  
The explanation of the bias when $N_o \leq N_c$ is technical and we remand to \citet{vio08}. When noise is present,
under the condition of $\Nmthb$ uncorrelated with $\Sb$, from model~(\ref{eq:model}) it is
\begin{equation} \label{eq:covar}
\Cb_{\Sb} = \Ab \Cb_{\Smthb} \Ab^T + \Omegab_{\Nmthb},
\end{equation}
with
\begin{equation}
\Omegab_{\Nmthb} = {\rm E}[ \Nmthb \Nmthb^T].
\end{equation}
Hence, a bias derives from the fact that $\Omegab_{\Nmthb}$ is a matrix with strictly positive entries and then
${\rm E}[\widehat \Cb_{\Sb}] \neq \Cb_{\Sb}$.

In absence of ``{\it a priori information}'', the problems connected to the first two points above have no solution. The only possibility
is to plan the experiments in such a way to avoid them (e.g. observations at high Galactic latitudes, a number of observing frequencies
sufficiently large \ldots). On the other hand, noise is an unavoidable question that, however, can be handled with hope of satisfactory
results. In this respect, some authors \citep[e.g. see][]{teg03, kim07, kim08} suggest that an effective way is to use ILC 
in the harmonic (Fourier) domain. In practice, this means to apply ILC to $\Shb$ that is the matrix whose $i$th row contains the two-dimensional 
Fourier transform of $\Sib$. Following \citet{kim07}, we will indicate this version of ILC as ``{\it harmonic internal linear combination}''
(HILC).

\section{ILC in the Fourier domain (HILC)} \label{sec:fourier}

The starting consideration is that all arguments presented in the previous sections hold independently of the fact that one is working
in the ordinary spatial domain or in the Fourier domain. Indeed, it is not difficult to see that the weights $\wb$ as given by Eq.~(\ref{eq:wr}) 
can be equivalently computed by means of
\begin{equation}
\wb = \frac{\Cb_{\Shb}^{-1} \oneb}{\oneb^T \Cb_{\Shb}^{-1} \oneb}.
\end{equation}
In this respect, it is sufficient to take 
into account that
\begin{equation}
\Shb = \Sb \Fmatb,
\end{equation}
with $\Fmatb = \Fb_{N_c} \otimes \Fb_{N_o}$, ``$\otimes$'' the Kronecker product,
$\Fb_{N}$ the $N \times N$ the Fourier matrix that is a complex, symmetric and
unitary matrix whose elements are given by
\begin{equation}
(F_{N})_{kl} = \frac{1}{\sqrt{N}} {\rm e}^{-2 \pi \iota (k-1)(l-1) /N},
\end{equation}
$\iota = \sqrt{-1}$. Hence
\begin{equation}
\Cb_{\Shb} = {\rm E}[\Shb \Shb^H] = {\rm E}[\Sb \Fb \Fb^H \Sb^T] = \Cb_{\Sb},
\end{equation}
since $\Fb \Fb^H = \Fb^H \Fb = \Ib$ where $\Fb_N^H$ is the complex conjugate transpose of $\Fb_N$.

\begin{figure*}
        \resizebox{\hsize}{!}{\includegraphics{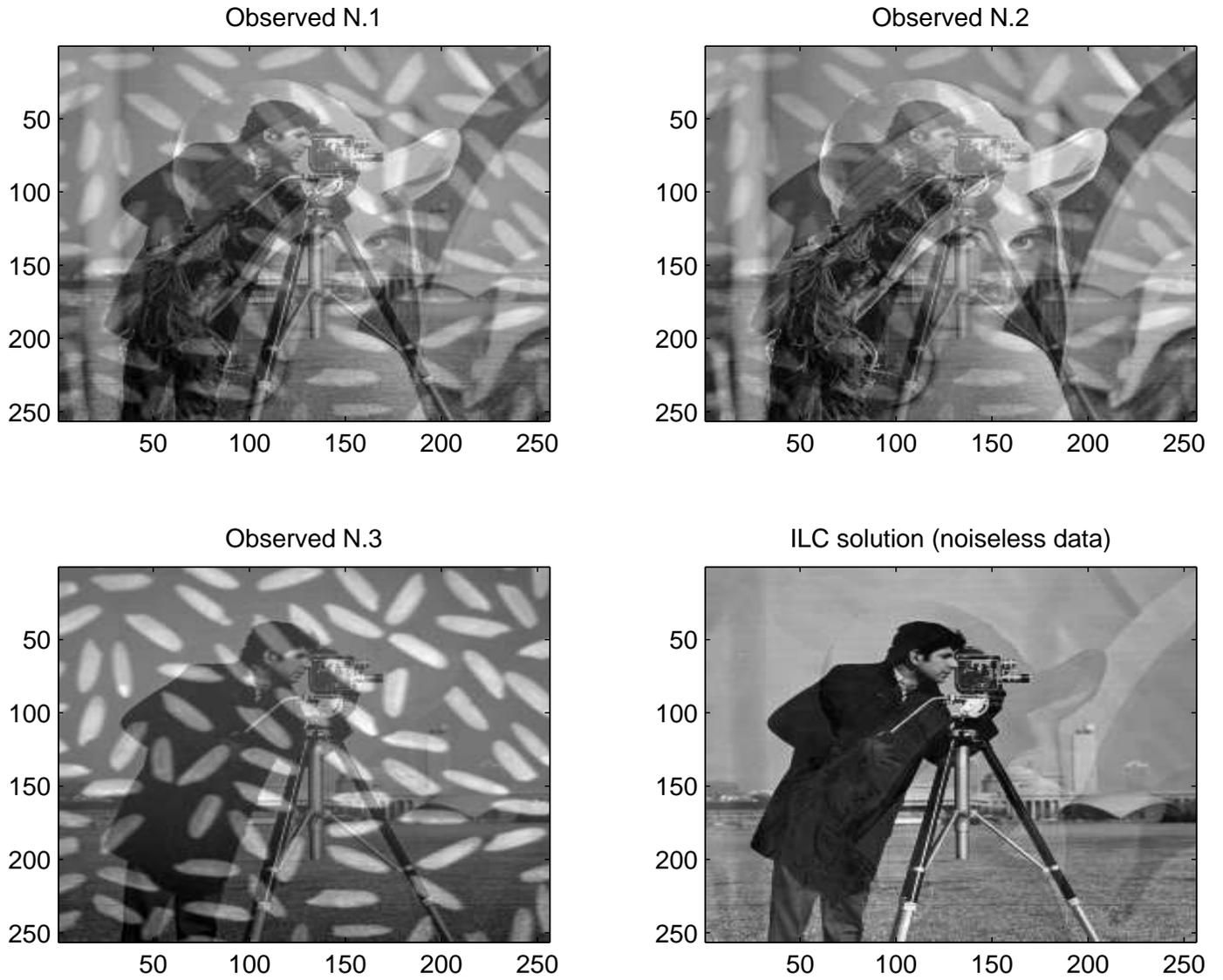}}
        \caption{Top panels and bottom-left panel: observed mixtures $\Sb^{(1)}$, $\Sb^{(2)}$ and $\Sb^{(3)}$ -- see Eq.~(\ref{eq:model}) --
        used in the experiment described in the text. The data shown here are noiseless. The right-bottom panel shows the solution obtained
        when ILC is applied to these images.}
        \label{fig:exp2}
\end{figure*}

It is well known that measurement noise $\{ \Nmthib \}$, typically the realization of wide-band stochastic
processes, tends to uniformly spread in the Fourier domain. On the
other hand, the contribution of signals $\{ \Sib \}$ is typically concentrated in correspondence to the lowest Fourier frequencies. For example,
this is visible in Fig.~\ref{fig:mask} where the power spectra of three simulated mixtures are shown (see Fig.~\ref{fig:exp2}).
Here, non-astronomical signals have been used, but the same holds also for astronomical ones. 
This implies that, for the indices ``$p$'' corresponding the lowest frequencies, the contribution of noise to $\Shi$ is quite small.
Hence, the basic idea here is to partition the frequency domain in a number $N_s$ of subsets and to apply separately the ILC filter to each of them.
At the end of this procedure 
an estimate $\widetilde \Smthb_c \equiv \bigcup_{k=1}^{N_s} \widetilde \Smth_c(p_k)$ is obtained where 
$p_k = \{ p | p \in k {\text -th~frequency~subset} \}$.
Finally, $\widehat \Smthb_c$ can be recovered by the Fourier inversion of $\widetilde \Smthb_c$.

\begin{figure*}
        \resizebox{\hsize}{!}{\includegraphics{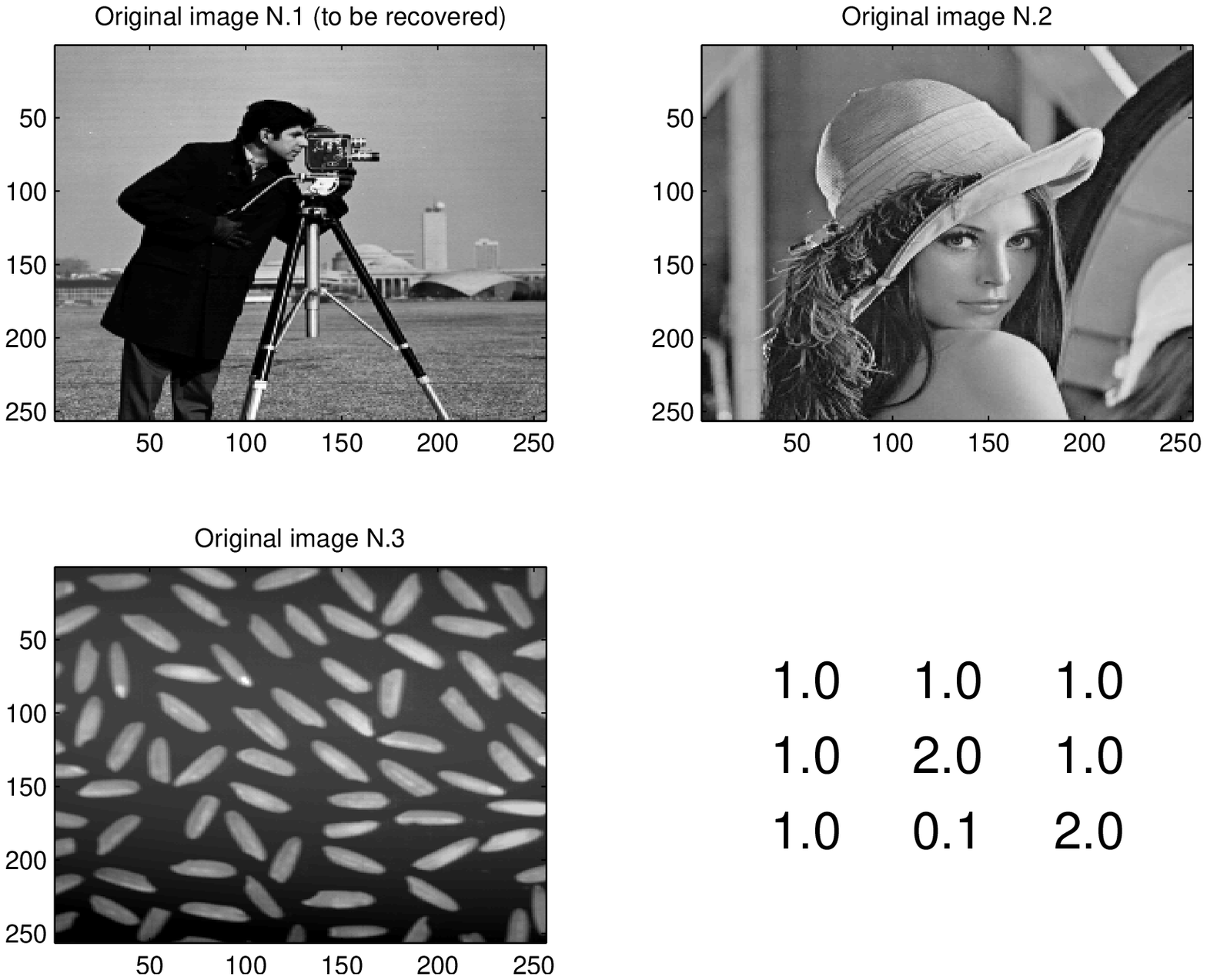}}
        \caption{Top panels and bottom-left panel: original images $\Smthb_c$, $\Smthb_1$ and $\Smthb_2$  -- see Eq.~(\ref{eq:modelS}) --
        used in the experiment described in the text. The top-left panel shows the image to recover. Through Eq.~(\ref{eq:model}) these
        images are used to produce the mixtures $\Sib$ shown in Fig.~\ref{fig:exp2}.
        The bottom-right panel provides the mixing matrix $\Ab$ -- see Eq.~(\ref{eq:Amatrix}).}
        \label{fig:exp1}
\end{figure*}

\begin{figure*}
        \resizebox{\hsize}{!}{\includegraphics{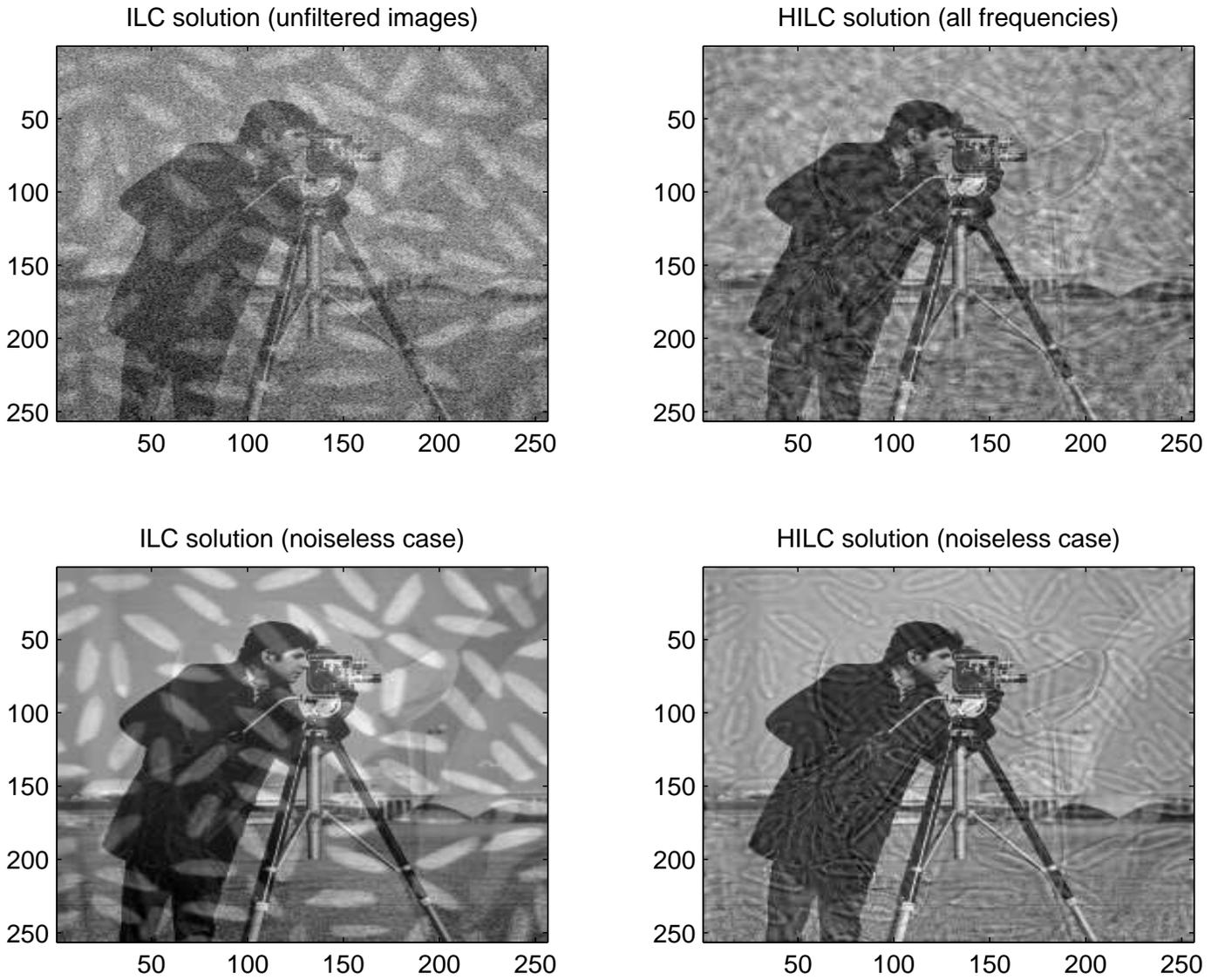}}
        \caption{Top panels: solutions provided by ILC (left panel) and HILC 
        (right panel) - see Sec.~\ref{sec:fourier} when a white-noise,
        with a signal-to-noise ratio set to $5$, is added to the images shown in Fig.~\ref{fig:exp2}. 
        Bottom panels: Results obtained when the weights used to compute 
        the solutions shown in the top panels are applied to the original (i.e. noise-free) images in Fig.~\ref{fig:exp2}).
        This operation gives an idea of the bias introduced in the solution by the noise.}
        \label{fig:exp3}
\end{figure*}

The effectiveness of such an approach seems supported by the simple example presented
in Figs.~\ref{fig:exp1}-\ref{fig:exp3}. In particular, Fig.~\ref{fig:exp1} shows the original images 
$\Smthb_c$, $\Smthb_1$, $\Smthb_2$ as well
the mixing matrix $\Ab$ that, through model~(\ref{eq:model}), are used to create the observed images $\{ \Sib \}_{i=1}^3$
shown in Fig.~\ref{fig:exp2}. As reference for later results, the bottom-right 
panel in the same figure shows the estimate $\widehat \Smthb_c$ obtained when ILC is applied to these images. 
For this case, the resulting {\it root mean square} (rms) of the residuals is about $0.05$. 
The top panels in Fig.~\ref{fig:exp3} show what happens when the images are added a white-noise with a {\it signal-to-noise} (SNR) 
\footnote{Here, the quantity {\rm SNR} is defined as the
ratio between the standard deviation of the signal with the standard deviation of the noise.} set to $5$. 
Here, both ILC and HILC are used. In the case of HILC, the frequency domain is 
partitioned in two regions, as shown in the bottom-left panel of Fig.~\ref{fig:mask}, one containing
the low frequencies and the other containing the high frequencies.
The weights $\wb$ are calculated independently for each of them. It 
is evident that HILC outperforms ILC. This conclusion is confirmed by the {\it rms} 
of the its residuals that is $0.13$ and $0.26$, respectively. The same indication is provided by the bottom panels that show
the result obtained when the weights computed for the noisy images are applied to the noise-free ones. This operation gives
an idea of the bias introduced in the solution by the noise. In this case, the {\it rms} of the residuals become $0.08$ and $0.18$,
respectively.

\section{Discussion and conclusions}

\begin{figure*}
        \resizebox{\hsize}{!}{\includegraphics{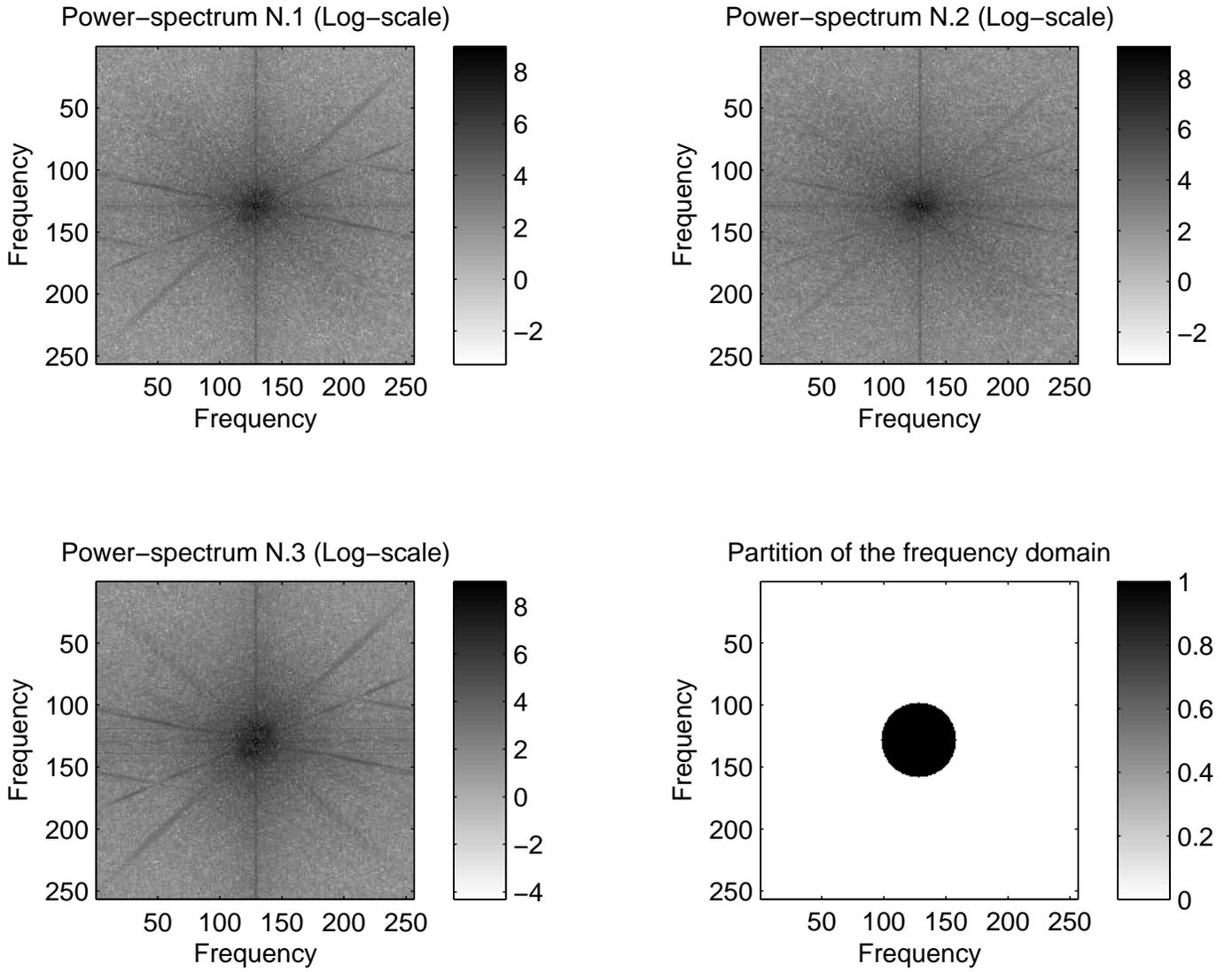}}
        \caption{Top panels and bottom-left panel: power spectra (logarithmic scale) of the images $\Sib$ shown in Fig.~\ref{fig:exp2}. 
        The Figure shows
        that most of the power is concentrated at low frequency. Bottom-right panel: partition of the discrete Fourier domain that 
        is used by HILC as discussed in Sect.~\ref{sec:fourier}.}
        \label{fig:mask}
\end{figure*}

\begin{figure*}
        \resizebox{\hsize}{!}{\includegraphics{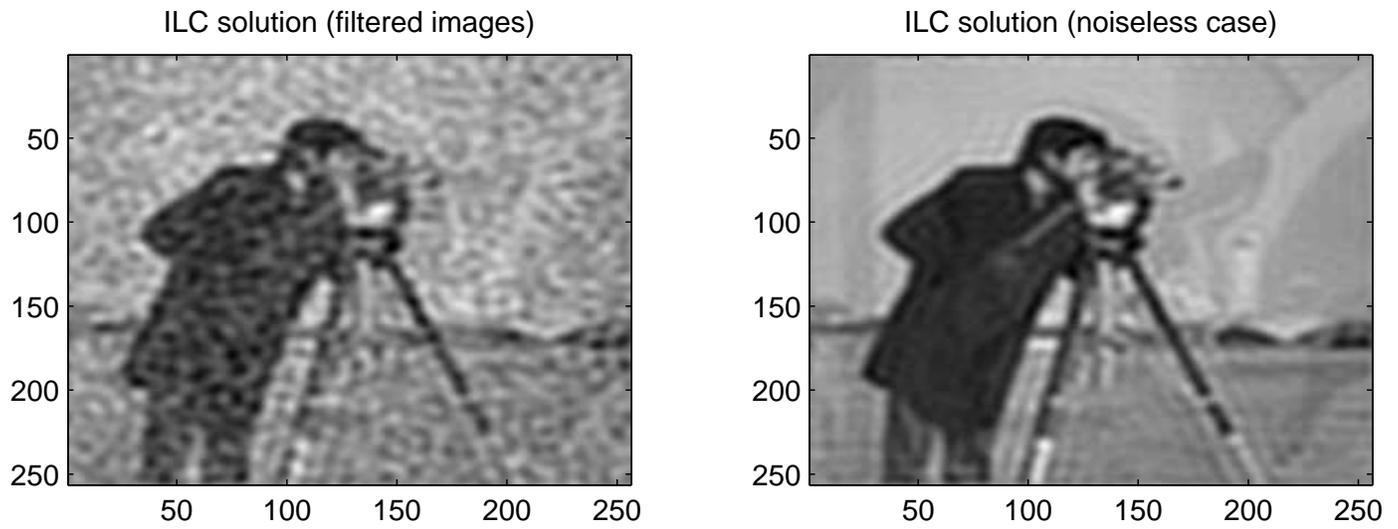}}
        \caption{As in the left panels of Fig.~\ref{fig:exp3} with the difference that before using ILC, the images have been
        filtered with the ideal circular low-pass filter that has the same structure as in the bottom-right panel of Fig.~\ref{fig:mask}.}
        \label{fig:exp4}
\end{figure*}

Reader should not be confused by the fact that at first sight HILC appears an approach potentially superior to ILC.
It is indeed true that HILC performs a more effective separation of the signal from the noise. However,
if the noise affects most the high frequency part of the signal, why then do not simply filter out this one?
Indeed, the left panel of Fig.~\ref{fig:exp4} shows the estimate 
$\widehat \Smthb_c$ obtained when, before the application of ILC, the images $\{ \Sib \}$ are filtered with 
the ideal Fourier circular low-pass filter \footnote{In the Fourier domain, an ideal filter $\widetilde \Hb$ has the 
form $\widetilde H(p) = 1$ if $p$ is a frequency of interest, $0$ otherwise.} that has the structure identical to that
shown in Fig.~\ref{fig:mask}. As before, the right panel of this figure shows the solution obtained when the resulting weights are applied to 
the noise-free version of the data. The {\it rms} of the residuals are $0.12$ and $0.08$, respectively. The comparison with the values
of $0.13$ and $0.08$, previously determined for HILC, indicates an improvement of the quality of the result.
The indication that comes out from this simple experiment is that filtering noise is a more effective operation than its separation from the signal of
interest. It is true that the version of HILC here used is rather rough and more sophisticated ones are available \citep{kim07, kim08}. 
However, the same holds also for the filtering
operation that has been coupled with ILC. At this point a question arises: which is the benefit in using a non-standard approach as HILC
with respect to a classic approach where filtering is coupled with ILC? This in not an academic question. Indeed, the use of non-standard tools implies
the renunciation of a huge body of experience gained in years of application in practical problems also in disciplines different 
from Astronomy. In other words, the choice of non-standard tools is indicated only in situations
of real and sensible improvements of the results. New techniques that do not fulfill this requirement should
be introduced with care: they prevent the comparison with the results obtained in other works and may lead
people to use not well tested methodologies ending up in not reliable results.

\clearpage

\end{document}